\documentclass[baaa]{baaa}

\usepackage[pdftex]{hyperref}
\usepackage{subfigure}
\usepackage{natbib}
\usepackage{helvet,soul, bm}
\usepackage[font=small]{caption}

\contriblanguage{1}

\contribtype{1}

\thematicarea{1}

\title{Observational evidence of fractality in the large-scale distribution of galaxies}

\titlerunning{Observational evidence of fractality in the large-scale distribution of galaxies}

\author{
T. Canavesi\inst{1,2}
\&
T.E. Tapia\inst{3}
}

\authorrunning{Canavesi \& Tapia}

\contact{tcanavesi@fisica.unlp.edu.ar}

\institute{
Insituto de F\'isica de la Plata, CONICET--UNLP, Argentina
\and
Facultad de Ciencias Astron\'omicas y Geof\'isicas, UNLP, Argentina
\and
Wolfram Research, EE.UU.
}

\resumen{
Usando una muestra de 133 991 galaxias distribuidas en la region del cielo $100^{\circ} <\alpha<270^{\circ}$ y $7^{\circ}<\delta<65^{\circ}$ con un redshift $0.0057<z<0.12$, extra\'ida del cat\'alogo SDSS NASA/AMES Value Added Galaxy Catalog ({\sc AMES-VAGC}), estimamos la dimensi\'on fractal usando dos m\'etodos. El primero usa un algoritmo para estimar la dimensi\'on de correlaci\'on. El segundo, una nueva aproximaci\'on, crea un grafo usando los datos y estima la dimensi\'on del grafo basado unicamente en la informaci\'on de conectividad. En ambos m\'etodos encontramos una dimensi\'on $D\approx 2$ en escalas por debajo de 20~\textrm{Mpc}, lo cual concuerda con trabajos previos. Este resultado muestra la no homogeneidad de la distribuci\'on de las galaxias en ciertas escalas.}

\abstract{
Using a sample of 133 991 galaxies distributed in the sky region $100^{\circ} <\alpha<270^{\circ}$ and $7^{\circ}<\delta<65^{\circ}$ with a redshift $0.0057<z<0.12$, extracted from the SDSS NASA/AMES Value Added Galaxy Catalog ({\sc AMES-VAGC}), we estimate the fractal dimension using two different methods. First, using an algorithm to estimate the correlation dimension. The second method, in a novel approach, creates a graph from the data and estimates the graph dimension purely from connectivity information. In both methods we found a dimension $D\approx 2$ in scales below 20~\textrm{Mpc}, which agrees with previous works. This result shows the non-homogeneity of galaxies distribution at certain scales.}


\begin{document}

\maketitle

\section{Context}\label{Context}

The cornerstone of modern cosmology is the cosmological principle, which assumes that the universe is homogeneous and isotropic at large scales. From this statement one could ask:
\begin{itemize}
    \item At what scale is the universe homogeneous? 
     \item Does the universe show a fractal structure?
\end{itemize}
There is no single definition for a fractal but we can say that a fractal often has some form of self-similarity either approximately or statistically. Specifically, a dimension $D=2$ in the distribution of galaxies corresponds with matter uniformly distributed on spherical surfaces surrounding the observation point. Several approaches to measure fractality have been proposed, \cite{chacon2016multi}, \cite{kamer2013barycentric}, \cite{chacon2009dinamica}. As a classical method, if we count the number of points $N$ inside a growing sphere of radius $r$, we expect for an  homogeneous distribution a power law relation of the form

\begin{equation}\label{number of points}
    N(r)\sim r^{D},
\end{equation}
with the dimension $D=3$. Observations in the last decades have allowed us making more accurate calculations of the dimension of galaxy distribution, as well as any other cosmological parameter. This work estimates the dimension of the spatial distribution of a sample of 133 991 galaxies extracted from the SDSS NASA/AMES Value Added Galaxy Catalog ({\sc AMES-VAGC})\footnote{\url{https://cdsarc.unistra.fr/viz-bin/cat/J/ApJ/799/95}} using two approaches. First, using the definition of correlation dimension. Second, in a novel way, by creating a graph from the galaxy spatial distribution and estimating its dimension purely from connectivity information. The catalog provides the cartesian positions of the galaxies in redshift units, which were transform to \textrm{Mpc} in the small redshift aproximation assuming assuming a value of $h=0.6767$ in the Hubble constant for simplicity.

\section{Methodology}\label{Methodology}

In the first approach we use the correlation integral. In agreement with \cite{bagla2008fractal}, the correlation integral $C_2$ is defined as

\begin{equation}\label{correlationint}
    C_{2}(r)=\frac{1}{N M}\sum_{i=1}^{M}n_{i}(r),
\end{equation}
where $N$ is the number of galaxies in the sample, $M$ is the number of galaxies chosen as centers of the growing spheres, n  is the number of galaxies reached by the growing spheres of radius $r$ with center in the $i$th-galaxy, and the summation is carried over the set of spheres. The correlation integral is defined similarly to \eqref{number of points}, then the correlation dimension is

\begin{equation}
    D_{2}=\frac{d \log C_{2}(r)}{d \log r}.
\end{equation}
In our results the growing spheres moves in steps of $2~\mathrm{Mpc}$. Then, the numeric estimation of the dimension is calculated as:

\begin{equation}\label{correlation integral}
D_2(r)=\frac{\log C_2(r+1)-\log C_2(r)}{\log(r+1)-\log(r)},
\end{equation}
where r now moves in discrete steps.

In the second method a graph is constructed from the position of the galaxies. Each galaxy represents a node, and an undirected edge will join two nodes if the distance between the nodes is less than 10~\textrm{Mpc}. This criterion in the distance is suitable because if it is smaller we obtain a graph with plenty of disconnected parts, and if it is greater we obtain an all connected graph. Then, to estimate the dimension of the resulting graph we use growing spheres whose center is a node $X$ in the graph, and we count the number of nodes we have inside each sphere until we reach the boundary of the graph. If we make an adjustment taking into account the number of nodes reached by each sphere as a function of the radius of the sphere, we can adjust a dimension $D$. This approach is similar to our first method, but in this case we just use connectivity information. For more details see chapter 4.5 of \cite{wolfram2020class}. 

The following computations were done using {\sc Mathematica 12.1} \footnote{\url{https://www.wolfram.com/mathematica/}}, and the software developed in the Wolfram Physics Project \footnote{\url{https://www.wolframphysics.org/}}.

\section{Results}\label{Results}

\begin{figure}[h]
\centering
\includegraphics[width=0.9\columnwidth]{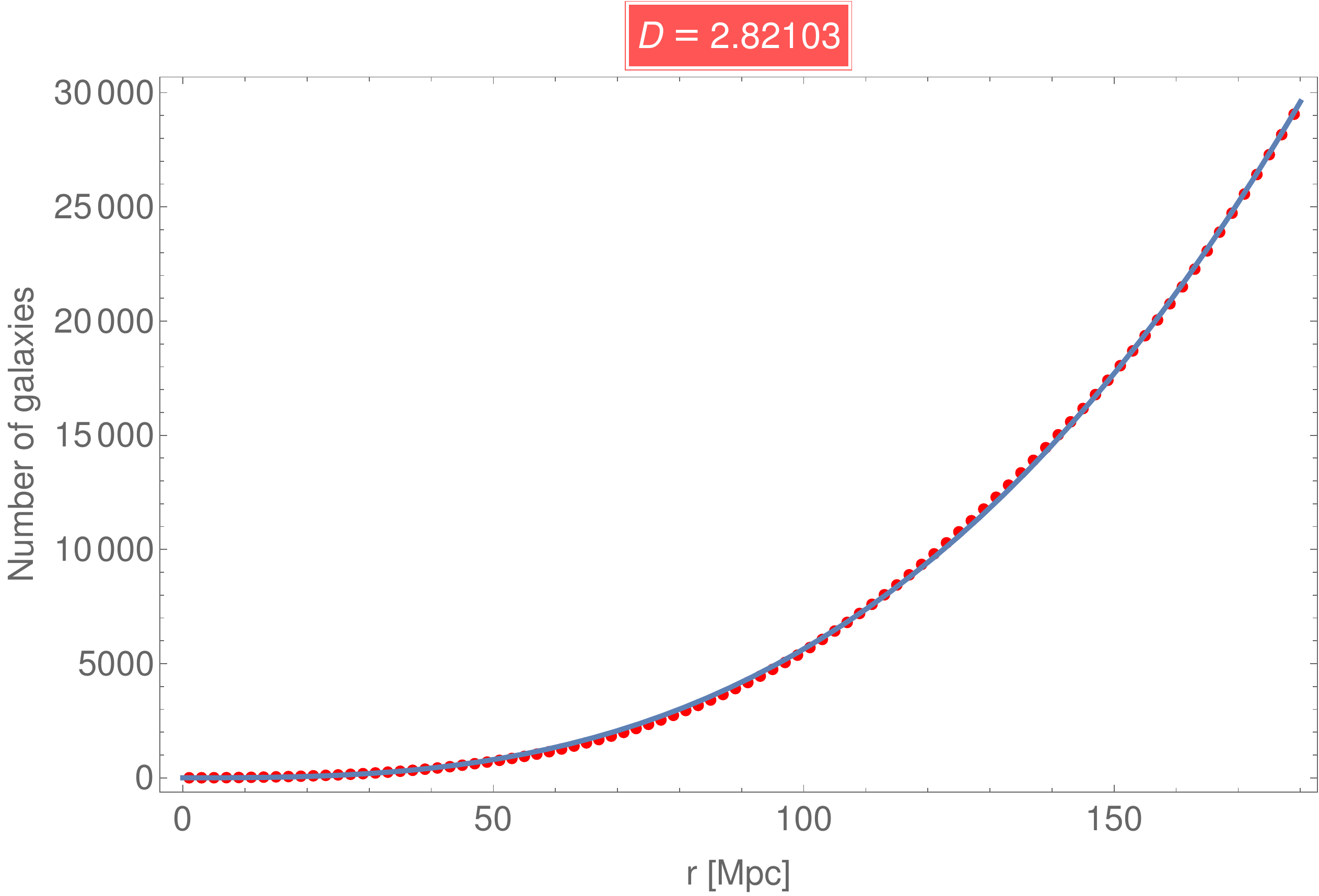}
\caption{Mean number of galaxies reached by the growing spheres vs the radius of the spheres. One hundred galaxies were used as centers of the growing spheres. Dots are obtained from the computation, and solid red line indicates a fit of the form $a r^D$. From the fit, the obtained dimension is shown in the red box.}
\label{Figura1}
\end{figure}

\begin{figure}[h]
\centering
\includegraphics[width=\columnwidth]{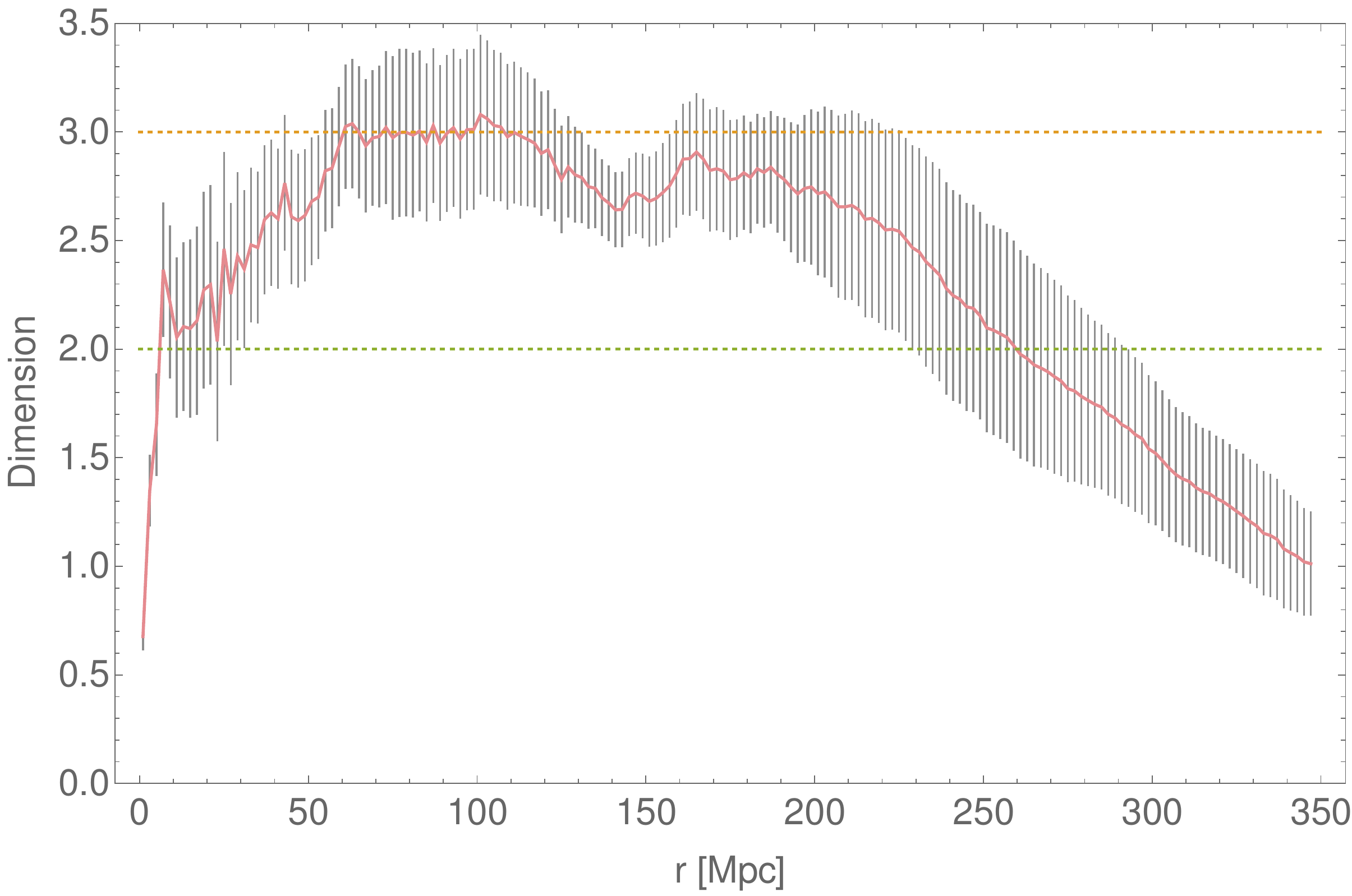}
\caption{Dimension vs. radius. Dimension is plotted along with gray error bars. Dashed lines indicate dimensions 2 and 3.}
\label{Figura2}
\end{figure}

Regarding the first method, Fig.~\ref{Figura1} shows the mean number of galaxies reached by the growing spheres vs the radius of the spheres. One hundred galaxies were used as centers of the growing spheres because with this number of centers local effects are not dominant in the average count. Error bars considering the $1\sigma$ ranges in the distribution of values were calculated.  Though not large enough to be significant in Fig.~\ref{Figura1}, their propagation imply noticeable uncertainties in  Fig.~\ref{Figura2}. Also, Fig.~\ref{Figura1} shows a fit of the form $a r^D$ (the red line), and the parameters of the fit are shown in Table.~\ref{tabla1}. A dimension $D=2.82103$ is obtained from the fit.   

\begin{table}[h]
\centering
\caption{Parameters of the fit  plotted in Fig.~\ref{Figura1}, including the standard error and the t-statistic.}
\begin{tabular}{lccc}
\hline\hline\noalign{\smallskip}
\!\!Parameter & \!\!\!\!Estimate & \!\!\!\!Standard Error& \!\!\!\!t-Statistic\!\!\!\!\\
\hline\noalign{\smallskip}
\!\!$a$ & 0.013 &0.00044 & 29.45 \\
\!\!$D$ & 2.82103& 0.00673 & 419.176\\
\hline
\end{tabular}
\label{tabla1}
\end{table}

\noindent Fig.~\ref{Figura2} shows the dimension $D$ as a function of the radius $r$. The decay of the dimension at large scales is due to boundary effects of the dataset. A dimension of $D=2$ in the scales of $[20, 30]~\mathrm{Mpc}$ transitions and reaches $D=3$ at the scales $[60, 70]~\mathrm{Mpc}$.

\begin{figure}[h]
\centering
\includegraphics[width=\columnwidth]{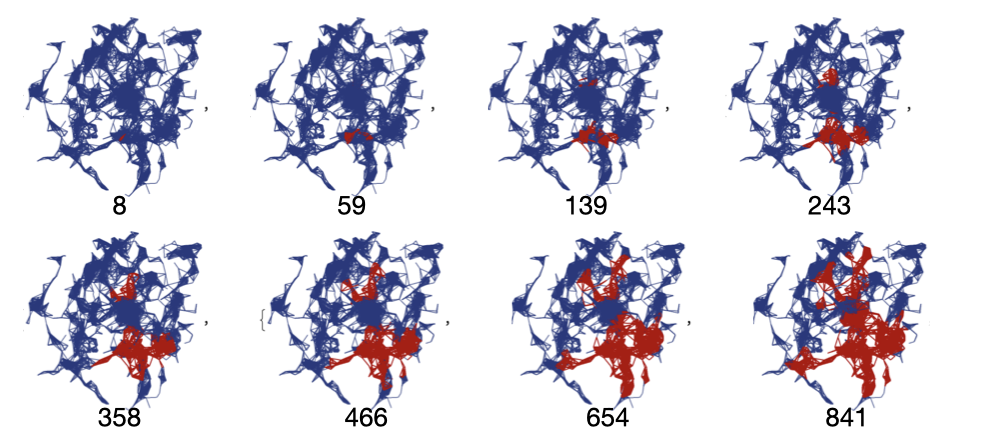}
\caption{A representative graph using 1853  nodes. Starting from a central node $X$ in the graph, we count the number of nodes in the graph that can be reached by going out at most a distance $r$. In red color we can observe this process starting from the center of the graph. The number indicated below each graph is the total number of nodes counted up to a distance $r$.}
\label{Figura3}
\end{figure}

\begin{figure}[h]
\centering
\includegraphics[width=0.85\columnwidth]{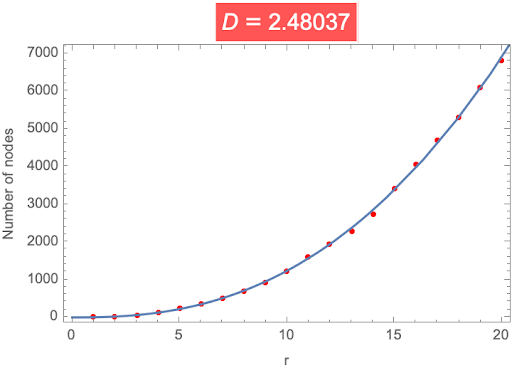}
\caption{Estimation of the dimension starting for a unique position in the graph.}
\label{Figura4}
\end{figure}

\begin{figure}[h]
\centering
\includegraphics[width=0.9\columnwidth]{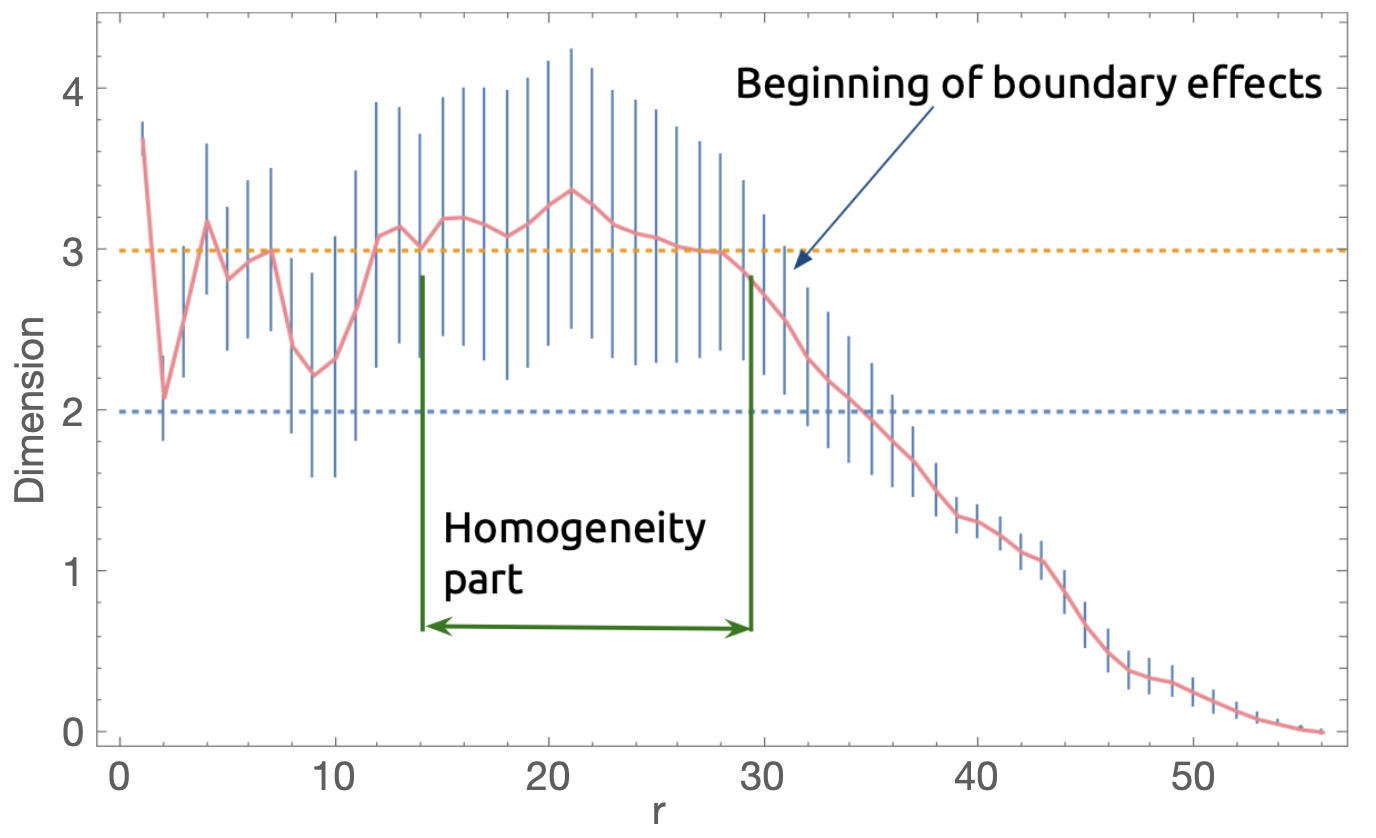}
\caption{Dimension estimation as a function of the distance $r$ from the central nodes $X$. The graph has 51 845 nodes and several central nodes $X$ are used. The error bars indicate $1\sigma$ ranges in the distribution of values obtained from different central nodes $X$.}
\label{Figura5}
\end{figure}

In our second method, after building a graph using the galaxies as explained in Sec.~\ref{Methodology}, we can estimate the dimension of the graph counting the number of nodes that fall within a sphere of increasing radius and whose limit is the boundary of the graph. We can see this process in a qualitative way in in Fig.~\ref{Figura3}.
In Fig.~\ref{Figura4} we can see the estimation of the dimension starting for a single point in the graph, in this case the center and using a fit of the form $a r^D$ whose adjustment parameters can be found in the in Table~\ref{tabla2}. In Fig.~\ref{Figura5} we show the dimension estimation using a graph of 51 845 nodes and considering 20 central nodes. We use several central nodes because the result may depend on certain central node $X$, then we average the counting using 20 central nodes. The error bars indicate $1 \sigma$ ranges in the distribution of values obtained from different central nodes $X$.

\begin{table}[h]
\centering
\caption{Parameters of the fit plotted in Fig.~\ref{Figura4}, including the standard error and the t-statistic.}
\begin{tabular}{lccc}
\hline\hline\noalign{\smallskip}
\!\!Parameter & \!\!\!\!Estimate & \!\!\!\!Standard Error& \!\!\!\!t-Statistic\!\!\!\!\\
\hline\noalign{\smallskip}
\!\!$a$ & 4.098 &0.256 & 15.974 \\
\!\!$D$ & 2.48037& 0.0219 & 113.182\\
\hline
\end{tabular}
\label{tabla2}
\end{table}

\section{Conclusions}\label{Concl}
\begin{itemize}
    \item Using the two methods, either the integral correlation or the graph approximation, a fractal dimension is found before the transition to homogeneity.
    \item Using the first approach, a transition from $D=2$ to $D=3$ starts in the interval $[20, 30]~\mathrm{Mpc}$ and ends in the interval $[60, 70]~\mathrm{Mpc}$. Other works report a transition to homogeneity at larger scales, as \cite{ntelis2017exploring}, \cite{scrimgeour2012wigglez}.
    \item Using the graph approach we found a region where the dimension of the graph is around $D=3$, and a region where the graph dimension is less than three, as seen with the first method.
     \item  We add evidence about the non-homogeneity of the distribution of galaxies at certain scales. Finding a transition from $D=2$, corresponding to a distribution of galaxies where the matter is uniformly distributed on spherical surfaces surrounding the observation point, to $D=3$ corresponding to a homogeneous and isotropic distribution of galaxies.
\end{itemize}
\subsection{Future Work}
\begin{itemize}
    \item Calculate the dimension of galaxy distribution using different methods.
    \item Use recent astronomical catalogs to consider more galaxies in the computations.
    \item Study the implications of fractality in the dynamics of galaxies using general relativity.
    \item Apply the methods used in this work to star formation regions, and compare with previous results as in \cite{2020BAAA...61B.254C}.
    \item Analyze the implications of assuming a fractal gravitational model in astronomical and cosmological scales, as in \cite{2020BAAA...61B.136C}. 
    
\end{itemize}

\begin{acknowledgement}
Thanks to the Wolfram Physics Project team for providing us with all the necessary software tools.
\end{acknowledgement}


\bibliographystyle{baaa}
\small
\bibliography{bibliografia}

\end{document}